\documentclass[11pt,twocolumn]{article}

\usepackage[a4paper,margin=2cm]{geometry}
\usepackage{graphicx}
\graphicspath{{../}{../figures/}}
\usepackage{amsmath,amssymb}
\usepackage{siunitx}
\usepackage[version=4]{mhchem}
\usepackage{booktabs}
\usepackage[numbers,sort&compress]{natbib}
\usepackage[hidelinks]{hyperref}
\usepackage{xcolor}
\usepackage{authblk}

\title{\bfseries Tunable Light Scattering in Cast Organic Scintillators via
\ce{BaSO4} Nanoparticle Doping: A Short Summary}

\author[1]{H.~G.~Zaunick,\footnote{Corresponding author (zaunick@exp2.physik.uni-giessen.de)}}
\author[1]{R.~Bergert}
\author[1]{K.~T.~Brinkmann}
\author[1]{V.~Dormenev}
\author[3]{K.~Eichhorn}
\author[3]{J.~M.~Friedrich}
\author[1]{S.~Glennemeier-Marke}
\author[1]{D.~Kazlou}
\author[2]{H.~Lacker}
\author[3,4,5]{M.~J.~Losekamm}
\author[2]{C.~Scharf}
\author[2]{I.~M.~W\"ostheinrich}

\affil[1]{Institute of Experimental Physics II, Justus Liebig University Giessen, Germany}
\affil[2]{Faculty of Mathematics and Natural Sciences, Humboldt University of Berlin, Germany}
\affil[3]{School of Natural Sciences, Technical University of Munich, Garching, Germany}
\affil[4]{Excellence Cluster ORIGINS, Garching, Germany}
\affil[5]{Now at European Space Agency, Noordwijk, Netherlands}

\date{\today}

\begin{document}
\maketitle

\begin{abstract}
\noindent
We summarize the fabrication and optical characterization of small-format
($2\times2\times2~\mathrm{cm}^3$) cast organic scintillators based on Eljen
EJ-290 resin, doped with barium sulfate (\ce{BaSO4}) powder at mass fractions
from \qty{0}{\percent} to \qty{5}{\percent}. The goal is to tune the scattering
length of the scintillator largely independently of its absorption and light
output, so that scintillation light is localized on a controllable spatial
scale matched to the fiber pitch of wavelength-shifting (WLS) fiber readout.
The scattering length is found to decrease from \SI{6.05(8)}{\centi\meter} at
\qty{1}{\percent} to \SI{0.83(1)}{\centi\meter} at \qty{5}{\percent}, while the
absolute light yield falls by only about \qty{15}{\percent}. These results are a
first proof of concept that the photon-transport scale in cast scintillators can
be engineered on purpose, enabling position-sensitive scintillator tiles for
fiber-readout sampling calorimeters and large-area muon trackers. Full details
are given in the accompanying paper.
\end{abstract}

\section{Motivation and concept}
Scintillator plates read out by WLS fibers and silicon photomultipliers (SiPMs)
are an attractive, scalable, low-cost building block for the active layers of
sampling calorimeters and large-area muon trackers~\cite{CheapCal2025}. The
achievable position resolution is governed by the spatial scale over which
scintillation photons spread before being captured by a fiber. In an
unsegmented, weakly scattering plate this scale is set by the photon transport
length, which is large (tens of centimeters); reducing the fiber pitch below it
then yields diminishing returns, because each fiber collects light from a broad
region and the centroid is poorly constrained.

\begin{figure}[t]
  \centering
  \includegraphics[width=\columnwidth]{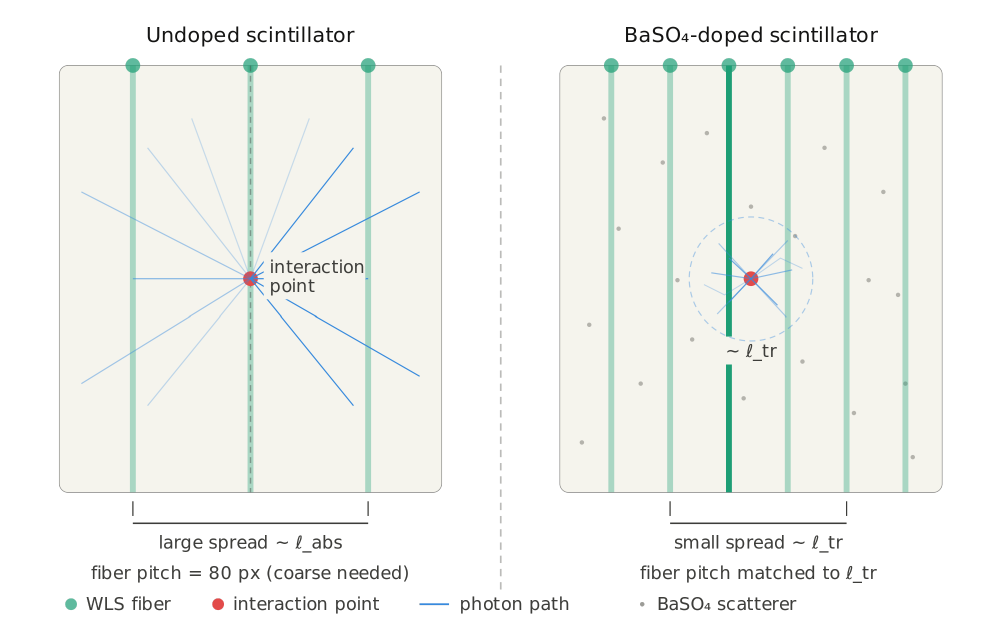}
  \caption{Light-localization concept. In an undoped scintillator (left) photons
  spread over a large area before capture; in a scattering-doped scintillator
  (right) the distribution is confined to a radius comparable to
  $\ell_\mathrm{tr}$, allowing the fiber pitch to be matched to this scale.}
  \label{fig:concept}
\end{figure}

Our idea is to deliberately shorten the transport mean free path
$\ell_\mathrm{tr}=\ell_\mathrm{s}/(1-g)$ by dispersing diffuse scattering
centers in the scintillator bulk, thereby localizing the light around the
particle interaction point (Fig.~\ref{fig:concept}).
In this case $\ell_\mathrm{s}$ is the scattering
length and $g$ the scattering anisotropy.
A finer fiber pitch matched to the reduced $\ell_\mathrm{tr}$ then translates directly into improved
position resolution. \ce{BaSO4} is a natural choice: it is a chemically inert,
high-reflectivity diffuse reflectance standard~\cite{Poh2019} with a low
refractive-index contrast to the organic host, so that scattering is nearly
isotropic ($g\approx0$, hence $\ell_\mathrm{tr}\approx\ell_\mathrm{s}$) and adds
negligible absorption~\cite{Krauter2015}.

\section{Sample fabrication}
\begin{figure}[t]
  \centering
  \includegraphics[width=\columnwidth]{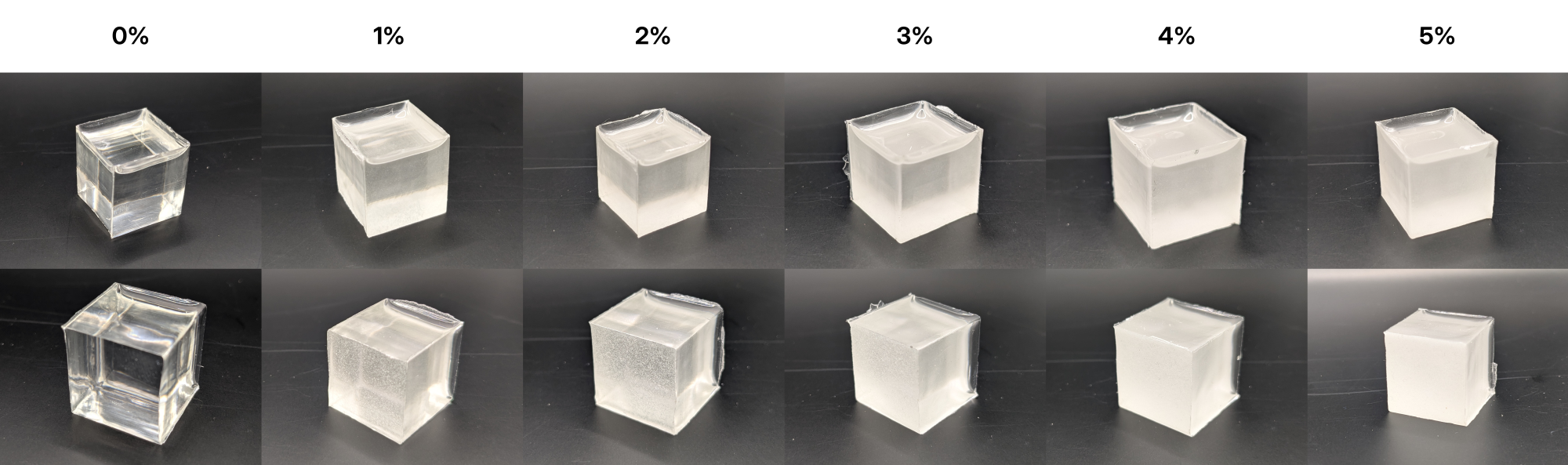}
  \caption{Cast samples without (left) and with \ce{BaSO4} at increasing
  concentration (weight percent), showing increasing opaqueness, i.e.\
  decreasing scattering length.}
  \label{fig:cubes}
\end{figure}

A series of six cubic samples ($2\times2\times2~\mathrm{cm}^3$) was cast from the
partially polymerized PVT-based EJ-290 resin~\cite{EJ290}. The manufacturer's
volumetric recipe was reformulated gravimetrically to allow accurate small-batch
mixing; \ce{BaSO4} powder (grain size $\sim$\SIrange{0.3}{2}{\micro\meter}) was
weighed in to the desired mass fraction and dispersed mechanically before adding
the hardener. The mixtures were degassed under vacuum, cast into custom
three-piece molds 3D-printed in PLA and lined with PTFE tape, and cured at
\qty{47}{\celsius} for two weeks. For optical characterization, two opposing
faces of each cube were flattened by diamond-tool milling. The undoped control
sample was optically clear; the samples grow visibly more opaque with increasing
\ce{BaSO4} content (Fig.~\ref{fig:cubes}). Some \ce{BaSO4} sedimentation was
still observed, indicating that the dispersion can be further improved.

\section{Characterization and results}
Spectral transmittance was measured with a spectrophotometer and converted, with
a Fresnel-reflection correction, to the attenuation length $\ell_\mathrm{att}$.
The absolute light yield (LY) was measured against a calibrated reference PMT
using a $^{241}$Am \SI{59.5}{\kilo\electronvolt} line. Taking the undoped sample
as a purely absorbing reference, the scattering length of each doped sample is
obtained from
$\ell_\mathrm{s}=1/(\mu^{(c)}-\mu^{(0)})$,
where $\mu=1/\ell_\mathrm{att}$ is the total attenuation coefficient. Values
extracted in the \SIrange{400}{430}{\nano\meter} band, near the scintillation
maximum (\SI{423}{\nano\meter}), are listed in Table~\ref{tab:results}.

\begin{table}[t]
  \centering
  \small
  \setlength{\tabcolsep}{4pt}
  \begin{tabular}{cS[table-format=4.0]S[table-format=2.2]S[table-format=1.3]S[table-format=1.2]}
    \toprule
    \ce{BaSO4} & {LY@\SI{4}{\micro\second}} & {$T$@\SI{423}{\nano\meter}} & {$\ell_\mathrm{att}$} & {$\ell_\mathrm{s}$}\\
    (wt.\%) & {(ph/MeV)} & {(\%)} & {(cm)} & {(cm)}\\
    \midrule
    0 & 9052 & 48.48 & 2.741 & {---}\\
    1 & 8801 & 31.23 & 1.886 & 6.05\\
    2 & 8273 & 25.75 & 1.596 & 3.82\\
    3 & 8541 & 19.62 & 1.311 & 2.51\\
    4 & 8199 &  6.91 & 0.779 & 1.09\\
    5 & 7632 &  3.92 & 0.638 & 0.83\\
    \bottomrule
  \end{tabular}
  \caption{Light yield, transmittance, attenuation length and derived scattering
  length versus \ce{BaSO4} concentration (statistical uncertainties on the last
  digit(s) are quoted in the full paper).}
  \label{tab:results}
\end{table}

The scattering length decreases monotonically by nearly an order of magnitude
across the studied range, from \SI{6.05(8)}{\centi\meter} at \qty{1}{\percent} to
\SI{0.83(1)}{\centi\meter} at \qty{5}{\percent} \ce{BaSO4}, while the absolute
light yield drops only mildly, from ca.~\SI{9000}{} to \SI{7600}{ph\per
\mega\electronvolt} (about \qty{15}{\percent}). The scattering scale can thus be
set largely independently of the scintillation output, over a millimeter-to-centimeter range well matched to practical WLS-fiber pitches; for reference, the
detector of Ref.~\cite{CheapCal2025} reaches a two-dimensional position
resolution below \SI{6}{\milli\meter} at a \SI{15}{\milli\meter} pitch. The main
systematic limitations --- the assumption of a concentration-independent
absorption coefficient and the limited dynamic range set by the
\SI{2}{\centi\meter} sample thickness --- are discussed in the full paper.

\section{Conclusion}
We have demonstrated that the scattering length of cast organic scintillators
can be controllably and reproducibly tuned by \ce{BaSO4} doping while largely
preserving the light output. This is a first proof of concept for engineering
the photon-transport scale on purpose, and it lays the groundwork for
position-sensitive scintillator tiles in which the fiber pitch is matched to a
tuned scattering length. Next steps are improving the homogeneity of the dispersion, e.g.\ via sonication, stabilizers, agitation during curing, or a controlled curing atmosphere. We also plan to explore further scatterers and to tune the spectral reflectance by admixing \ce{SiO2} nanoparticles~\cite{mikhailov2018}. Finally, we aim to demonstrate the concept in a full detector plate with fiber readout.
The approach is complementary to other opaque-scintillator developments such as
wax-based and water-based opaque media~\cite{Buck2019,Apilluelo2025}.

\paragraph{Acknowledgments.}
This work was funded by the German Federal Ministry of Education and Research
(BMFTR) within the High-D-Calo consortium (05H2024), grant 05H24RGA.

\bibliographystyle{unsrtnat}
\bibliography{cheapcal}

\end{document}